\documentclass[aps,groupedaddress,superscriptaddress,onecolumn,prl,preprint]{revtex4}
\usepackage{amsmath}
\usepackage{amssymb}
\usepackage{graphicx}
\usepackage{amsthm}
\usepackage{hyperref}
\usepackage{bbold}
\usepackage{xcolor}
\usepackage{longtable}
\usepackage{booktabs}

\begin{document}

\newcommand*{\cl}[1]{{\mathcal{#1}}}
\newcommand*{\bb}[1]{{\mathbb{#1}}}
\newcommand{\ket}[1]{|#1\rangle}
\newcommand{\bra}[1]{\langle#1|}
\newcommand{\inn}[2]{\langle#1|#2\rangle}
\newcommand{\proj}[2]{| #1 \rangle\!\langle #2 |}
\newcommand*{\tn}[1]{{\textnormal{#1}}}
\newcommand*{\1}{{\mathbb{1}}}
\newcommand{\T}{\mbox{$\textnormal{Tr}$}}
\newcommand{\todo}[1]{\textcolor[rgb]{0.99,0.1,0.3}{#1}}

\theoremstyle{plain}
\newtheorem{prop}{Proposition}
\newtheorem{proposition}{Proposition}
\newtheorem{theorem}{Theorem}
\newtheorem{lemma}{Lemma}
\newtheorem{corollary}{Corollary}[proposition]
\newtheorem{remark}{Remark}

\theoremstyle{definition}
\newtheorem{definition}{Definition}

\title{\normalsize{Enhanced Scalability in Assessing Quantum Integer Factorization Performance}}

\author{Junseo Lee}
\email{harris.junseo@gmail.com}
\affiliation{Quantum AI Team, Norma Inc., Seoul 04799, Korea}

\date{\today}

\begin{abstract}
With the advancement of quantum technologies, there is a potential threat to traditional encryption systems based on integer factorization. Therefore, developing techniques for accurately measuring the performance of associated quantum algorithms is crucial, as it can provide insights into the practical feasibility from the current perspective. In this chapter, we aim to analyze the time required for integer factorization tasks using Shor's algorithm within a gate-based quantum circuit simulator of the matrix product state type. Additionally, we observe the impact of parameter pre-selection in Shor's algorithm. Specifically, this pre-selection is expected to increase the success rate of integer factorization by reducing the number of iterations and facilitating performance measurement under fixed conditions, thus enabling scalable performance evaluation even on real quantum hardware.
\end{abstract}

\maketitle


\section{Introduction}
Quantum information theory encompasses various topics such as quantum communication, quantum error correction, and quantum simulations. The ability to manipulate and control quantum systems has brought about the development of new technologies and the potential to solve problems that are computationally challenging for classical computers. Over the past few decades, there has been significant progress in quantum computing hardware and algorithms; however, a fully fault-tolerant quantum computer in its perfect form does not yet exist.

As computer systems evolve in both hardware and software aspects, the potential threat of quantum technologies in cybersecurity has been evident for several years. Therefore, understanding cryptographic algorithms from the perspective of quantum information theory has become a significant task in this field. It is well-known that traditional cryptographic algorithms such as the RSA cryptosystem, elliptic curve cryptography, Diffie-Hellman key exchange, and signature schemes are no longer considered secure with the emergence of integer factorization using Shor's algorithm~\cite{1} and searching using Grover's algorithm~\cite{2}. Here, we raise a practical question: How significant is the potential threat in the context of current quantum computing resources?

Given the potential of quantum information theory and the increasing development of quantum technologies, a variety of research studies on quantum information theory are presented to explore its potential effect in the cybersecurity field~\cite{3, 4, 5, 6, 7}. For example, research on Shor's algorithm~\cite{3, 6} explained its capability of breaking the widely used RSA public key cryptosystem developed by Rivest, Shamir, and Adleman~\cite{8}, which takes advantage of the inherent potential of quantum computers. The advent of Shor's algorithm attracted significant interest from the public, resulting in a thorough analysis of the commercialization of quantum computers. This is due to the algorithm's ability to efficiently factor large numbers, potentially breaking modern classical cryptography systems. In addition, several studies~\cite{4, 7} evaluated quantum attacks on the Advanced Encryption Standard (AES)~\cite{9} developed by NIST using the ability of an exhaustive key search technique by Grover's algorithm~\cite{2}.

With the substantial support from various literature sources, the performance of quantum computing algorithms is assessed under specific conditions. Typically, such evaluations occur in simulated quantum computing environments aimed at solving particular mathematical problems. One prominent challenge is Shor's algorithm, which is scrutinized for its potential cybersecurity implications, given its status as the most well-known method for integer factorization—an essential assumption underlying many encryption methods believed to be resistant to classical computing. However, previous studies are limited in their focus on solving specific values through tailored optimization approaches~\cite{10}. While case-by-case experiments shed light on the potential threat landscape, fully comprehending the scale of this threat using current quantum computing resources remains challenging from a broader perspective.

This study systematically evaluates the performance of integer factorization using Shor's algorithm in a gate-based quantum computing environment. To accomplish this, we conducted tests using the IBM quantum simulator, employing pre-selection of random parameters when performing integer factorization with Shor's algorithm. This pre-selection allows us to conduct scalable evaluations of Shor's algorithm across various ranges of $N$ (odd square-free semiprime numbers).

\section{Theorectical Background}
In this section, an overview of the background knowledge for this chapter is provided, such as Shor's algorithm, quantum Fourier transform, and matrix product state quantum simulation.

\subsection{Shor's algorithm}
Shor's algorithm is a well-known example of a quantum algorithm for factoring integers. The detailed steps of Shor's algorithm are as follows:
\begin{enumerate}
    \item Choose a random integer $a$ between $1$ and $N-1$.
    \item Compute the greatest common divisor (GCD) of $a$ and $N$ using the Euclidean algorithm.
    \item If GCD is not equal to $1$, then we have found a nontrivial factor of $N$ and we are done.
    \item If GCD is equal to $1$, use the quantum period-finding subroutine to find the period $r$ of the function $f(x) = a^x~\text{mod}~N$. (Please refer to Section~\ref{sec:order} for further details.)
    \item If $r$ is odd, go back to step 1 and choose a different $a$.
    \item If $a^{r/2} = -1~\text{mod}~N$, go back to step 1 and choose a different $a$.
    \item Otherwise, calculate the GCD of $a^{r/2} \pm 1$ and $N$. If either GCD is a nontrivial factor of $N$, we have successfully factored $N$.
\end{enumerate}

The most important step among them is the calculation of the smallest $r$, where we need to find the period $r$ (s.t. $f(r)=1$) of the function $f(x) = a^x~\text{mod}~N$ (step 4). To achieve this, we use a quantum period-finding subroutine. The quantum period-finding algorithm employs the quantum Fourier transform to efficiently find the period $r$ of the function $f(x)$. The quantum Fourier transform is the quantum analog of the classical Fourier transform and is a key component of various quantum algorithms. It enables the transformation of a quantum state encoding a function into a superposition of its periodic components.

\subsection{Quantum Fourier transform}
The quantum Fourier Transform (QFT) is a quantum algorithm that performs a Fourier transform on a quantum state. It is used in various quantum algorithms, including Shor's algorithm for factoring integers and Grover's algorithm for searching an unsorted database. The QFT is similar to the classical Fourier transform, but it uses quantum gates and operations to perform the transformation. The QFT takes a quantum state $|x\rangle$, where $x$ is a binary string of length $n$, and maps it to another quantum state $|y\rangle$, where $y$ is a binary string of length $n$. The transformation is defined as follows:

\begin{equation}
    \text{QFT}|x\rangle = \frac{1}{\sqrt{2^n}}\sum_{y=0}^{2^n-1} e^\frac{2\pi ixy}{2^n} |y\rangle
\end{equation}

The QFT can be implemented using a series of quantum gates, including the Hadamard gate, phase shift gates, and controlled-phase shift gates.

\subsection{Matrix product state quantum simulation}\label{sec:mps}
The IBM quantum circuit simulators are software tools that enable users to simulate the behavior of quantum computers on classical computers. These simulators are meticulously crafted to deliver precise and efficient simulations of quantum circuits, empowering researchers and developers to thoroughly test and optimize their algorithms prior to deploying them on actual quantum hardware. Table~\ref{tab:tab1} presents an overview of the currently supported IBM simulators, specifying the maximum number of qubits they can accommodate, as well as the theoretically breakable bits in the RSA scheme using Shor's algorithm.

\begin{table}
\caption{The types of IBM simulators and the theoretically breakable bits $\beta$ depend on their supported number of qubits $Q$}
\centering
\label{tab:tab1}
\begin{tabular}[t]{|c|c|c|}
\hline
\textbf{Quantum Simulator (Type)}      & \textbf{$Q$} & \textbf{$\beta$} \\ \hline\hline
Statevector simulator (Schr\"{o}dinger wavefunction)         & 32                         & 7                       \\ \hline
Stabilizer simulator (Clifford)           & 5000                       & 1249                    \\ \hline
Extended stabilizer simulator (Extended Clifford (e.g., Clifford+T)) & 63                         & 15                      \\ \hline
MPS simulator (Matrix Product State)                  & 100                        & 24                      \\ \hline
QASM simulator (General, context-aware)           & 32                         & 7                       \\ \hline
\end{tabular}
~
\end{table}

This study specifically utilized the Matrix Product State (MPS) simulator for simulating quantum circuits. With support for up to 100 qubits, it facilitated the evaluation of Shor's algorithm for integer factorization of numbers up to 24 bits. In comparison, the stabilizer simulator, which supports the most qubits, was limited to simulating Clifford gates and lacked support for non-Clifford gates essential for effectively implementing Shor's algorithm.

Especially, MPS simulations are known to be advantageous for quantum circuits with weak entanglement~\cite{11}. The fundamental element of the MPS simulation protocol entails locally decomposing the general pure state of the $n$ spins, denoted as $|\Psi\rangle\in \mathcal{H}_2^{\otimes n}$, into $n$ tensors $\{\Gamma^{[l]}\}_{l=1}^n$ and $n-1$ vectors $\{\lambda^{[l]}\}_{l=1}^{n-1}$ within the given two-dimensional Hilbert space $\mathcal{H}_2$. Here, the tensor $\Gamma^{[l]}$ is assigned to qubit $l$ and possesses at most three indices, denoted as $\Gamma^{[l]i}_{\alpha\alpha'}$, where $\alpha, \alpha'$ range from 1 to $\chi$ (here, $\chi$ is termed the \textit{Schmidt number}, signifying the maximum Schmidt rank across all conceivable bipartite partitions of the $n$ qubit system), and $i$ ranges from 0 to 1. Conversely, $\lambda^{[l]}$ represents a vector with its elements $\lambda^{[l]}_{\alpha'}$ storing the Schmidt coefficients of the partitioning $[1:l]:[(l+1):n]$. Accordingly, the quantum state of multiple particles across $N$ sites is represented by the following structure:
\begin{equation}
|\Psi\rangle = \sum_{i_1=0}^{1}\cdots\sum_{i_n=0}^{1} c_{i_1\cdots i_n} |i_1\rangle \otimes \cdots \otimes |i_n\rangle,
\end{equation}

where the coefficients $c_{i_1\cdots i_n}$ are computed as:
\begin{equation}
c_{i_1\cdots i_n} = \sum_{\alpha_1\cdots\alpha_{n-1}} \Gamma_{\alpha_1}^{[1]i_1}\lambda_{\alpha_1}^{[1]} \Gamma_{\alpha_1\alpha_2}^{[2]i_2} \lambda_{\alpha_2}^{[2]}\cdots\Gamma_{\alpha_{n-1}}^{[n]i_n},
\end{equation}

with the constraint $\sum_{i_1=0}^1\cdots\sum_{i_n=0}^1 |c_{i_1\cdots i_n}|^2 =1$.

The memory requirements for the computation increase proportionally to $\chi^2n$, while efficient simulation of quantum circuits in terms of memory and computational demands is feasible when $\chi \sim \text{poly}(n)$, signifying weak entanglement. Accordingly, MPS simulations can vary in computational and memory requirements depending on the Schmidt number between two qubits, and the structure of quantum circuits such as the order of quantum registers can significantly impact simulation time. Further analysis on this matter will be conducted in Section~\ref{sec:ent}.

\section{Integer Factorization using Selected Parameter}
\subsection{The Effect of Random Parameters}
Random parameters play a crucial role in the effectiveness of Shor's algorithm. Specifically, the selection of a random parameter $a$ in the first step is pivotal for the algorithm's efficiency. If a random number sharing a common factor with $N$ is chosen, the greatest common divisor (GCD) calculated in the second step will exceed 1, enabling the direct identification of a nontrivial factor of $N$. However, opting for a random number coprime to $N$ results in a GCD of 1, necessitating reliance on the quantum period-finding subroutine to determine the period $r$ of the function $f(x) = a^x~\text{mod}~N$. Hence, the algorithm's success hinges on the stochastic selection of parameters and the probabilistic nature of quantum measurement.

\begin{figure}[ht]
\centering
\includegraphics[width=16cm]{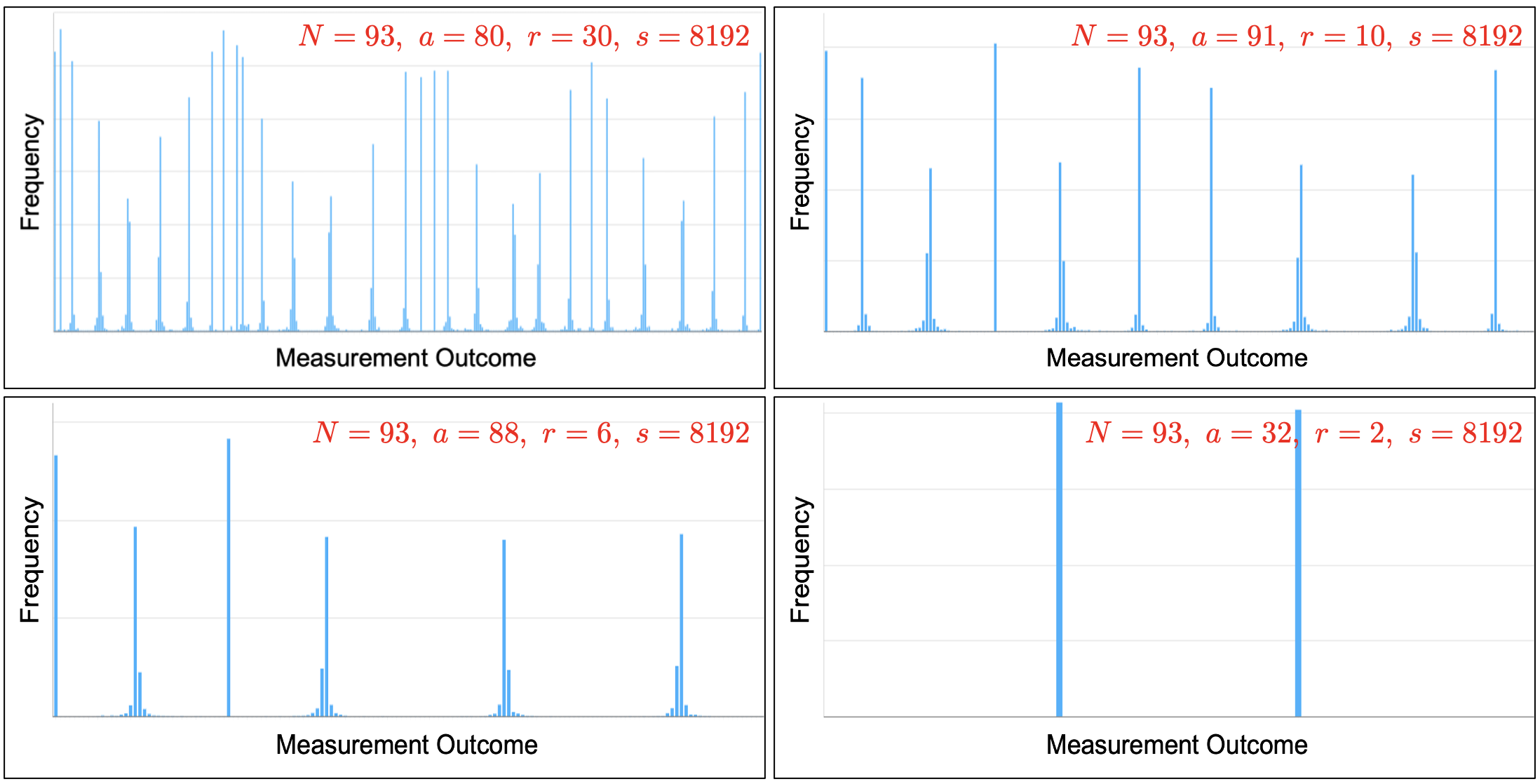}
\caption{Simulation results of the quantum period-finding subroutine in Shor’s algorithm when varying $N,~a,~r$}
\label{fig:fig1}
\end{figure}

The outcomes of the quantum period-finding subroutine in Shor's algorithm are depicted as probability distributions based on measurement outcomes, as illustrated in Figure~\ref{fig:fig1}. In an ideal quantum computing environment with a sufficient number of shots, the number of histogram bars would precisely correspond to the value of $r$. However, in the presence of noise, as observed in the figures, errors are introduced when employing a noisy quantum simulator or quantum hardware. Insufficient iterations can render the interpretation of experimental findings challenging. It is anticipated that larger values of $r$ will necessitate more iterations. For instance, when $r = 2$, only two bars appear in the histogram, allowing for the attainment of the quantum period-finding subroutine's outcome with a limited number of iterations. Figure~\ref{fig:fig1} exhibits the results obtained from factoring $N = 93$, with corresponding values of $a = 80$, $91$, $88$, and $32$, leading to respective $r$ values of $30$, $10$, $6$, and $2$. Here, $s$ denotes the number of shots for the quantum circuit.

Thus, we pre-select the value of $a$ with a minimum value of $r$ (which is $2$) and use it instead of randomly selecting $a$ for scalable testing of Shor's algorithm. However, this method also has limitations as $N$ increases. Therefore, we want to compare the time required to factor using a quantum simulator as $N$ increases while restricting $a$ to the same value of $r$. Table~\ref{tab:tab2} shows several values of $a$ that make $r$ equal to $2$ for actual semiprime $N$. To measure the performance of the simulator under standardized conditions, the $(N, a)$ pairs listed below are executed together instead of selecting a random $a$.

\begin{table}[]
\caption{Examples values of $a$ where $r = 2$}
\centering
\begin{tabular}[t]{c||c|c|c|c|c|c}
\hline
$N$ & 15 & 129 & 335 & 687 & 7617 & 9997 \\ \hline
$a$ & 4  & 44  & 66  & 230 & 2540 & 768 \\ \hline
\end{tabular}
\label{tab:tab2}
~
\end{table}

\subsection{Quantum order finding routine}\label{sec:order}
The quantum order finding routine in Shor's algorithm is implemented based on the following steps:
\begin{enumerate}
    \item \textit{Prepare the quantum state}: You first need to prepare a quantum state that is a superposition of all possible values of the period of the function $f(x) = a^x~\text{mod}~N$. This involves applying a Hadamard gate to a set of qubits that will represent the period.

    \item \textit{Apply the modular exponentiation function}: Apply the modular exponentiation function to the state, which maps the state $|x\rangle$ to $|a^x~\text{mod}~N\rangle$. This involves applying controlled gates that implement the function $f(x) = a^x~\text{mod}~N$. Specifically, for each qubit in the period register, you apply a controlled-$U_{a^{2^j}}~\text{mod}~N$ gate to the state, where $U_{a}$ is a unitary that performs the function $f(x) = a^x~\text{mod}~N$.

    \item \textit{Apply the inverse quantum Fourier transform}: Apply the inverse quantum Fourier transform to the period register, which maps the state $|r\rangle$ to a state that is proportional to $|\frac{r}{f}\rangle$, where $r$ is the period and $f$ is the order of the Fourier transform. This involves applying a set of controlled-phase gates that depend on the position of the qubits in the period register.

    \item \textit{Measure the period register}: Measure the period register and use classical post-processing to find the period. Specifically, you measure the period register and obtain a value $r$.
    
    \item \textit{Continued fractions algorithm}: Take the ratio $\frac{r}{f}$ and create a continued fraction $[a_0, a_1, a_2, a_3, \cdots]$. Determine the whole number component of the continued fraction as $a_0 = \lfloor\frac{r}{f}\rfloor$. Adjust the fraction to be $\frac{r}{f} = \left(\frac{r}{f} - a_0\right)^{-1}$. Determine the next whole number component $a_i$ as $a_i = \lfloor \left(\frac{r}{f} - a_{i-1}\right)^{-1} \rfloor$. Adjust the fraction to be $\left(\frac{r}{f} - a_{i-1}\right)^{-1} = a_i + \left(\frac{r}{f} - a_i\right)^{-1}$. Continue doing this until the fraction becomes a repeating sequence or until a maximum number of iterations is reached. The period of the continued fraction is the length of the repeating sequence. If the length of the repeating sequence is odd, repeat the last value in the sequence to make it even. The period of the continued fraction is the smallest value of $r$ that satisfies the equation $a^r~\text{mod}~N = 1$.
\end{enumerate}

\subsection{Entanglement Analysis for MPS}\label{sec:ent}
As discussed in Section~\ref{sec:mps}, the arrangement order of quantum registers is crucial for efficient simulation. In circuits implementing Shor's algorithm, registers are broadly categorized into \texttt{upper}, \texttt{lower} and \texttt{ancilla}. Therefore, by systematically changing their order and computing the von Neumann entropy, one can assess the degree of entanglement between one register and the rest. Figure~\ref{fig:fig2} below illustrates the experimental results corresponding to the case of $N=15$ and $a=4$. Analyzing these experimental results reveals that the entanglement between the \texttt{Upper} register and the combined \texttt{Lower+Ancilla} registers is consistently strong on average. Hence, for MPS simulation, it is appropriate to consider the order of \texttt{Upper-Lower-Ancilla} or \texttt{Upper-Ancilla-Lower}.
\begin{figure}[ht]
\centering
\includegraphics[width=16cm]{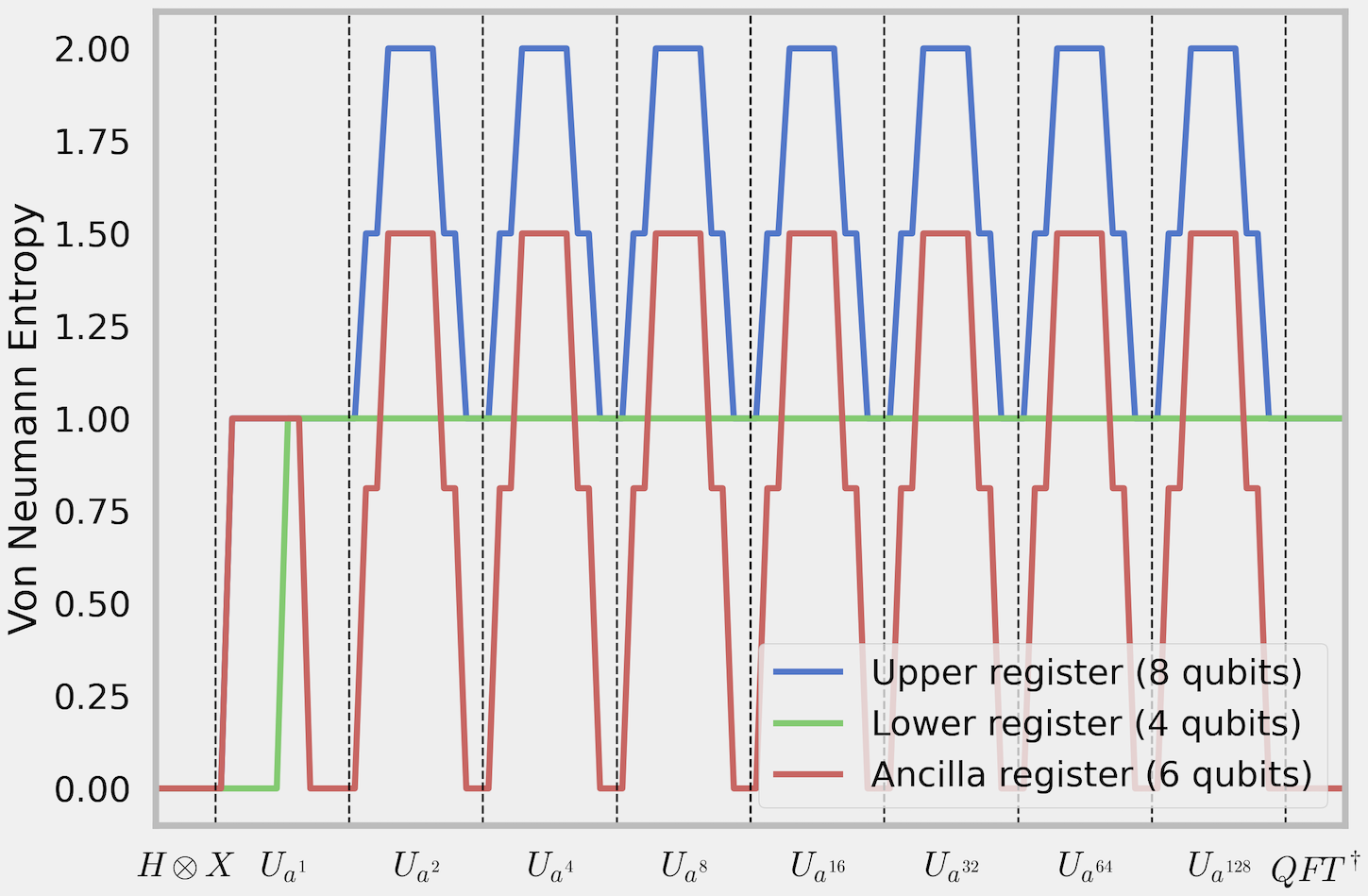}
\caption{Analysis of the degree of entanglement for each register}
\label{fig:fig2}
\end{figure}

\section{Performance of Shor's Algorithm at Scale}
In this section, the time taken for integer factorization using Shor's algorithm on the IBM quantum circuit simulator is presented under two scenarios: pre-selection of parameter $a$ and random selection of parameter $a$. All factorization times represent the duration required for the aforementioned quantum period-finding subroutine.

\begin{figure}[ht]
\centering
\includegraphics[width=16cm]{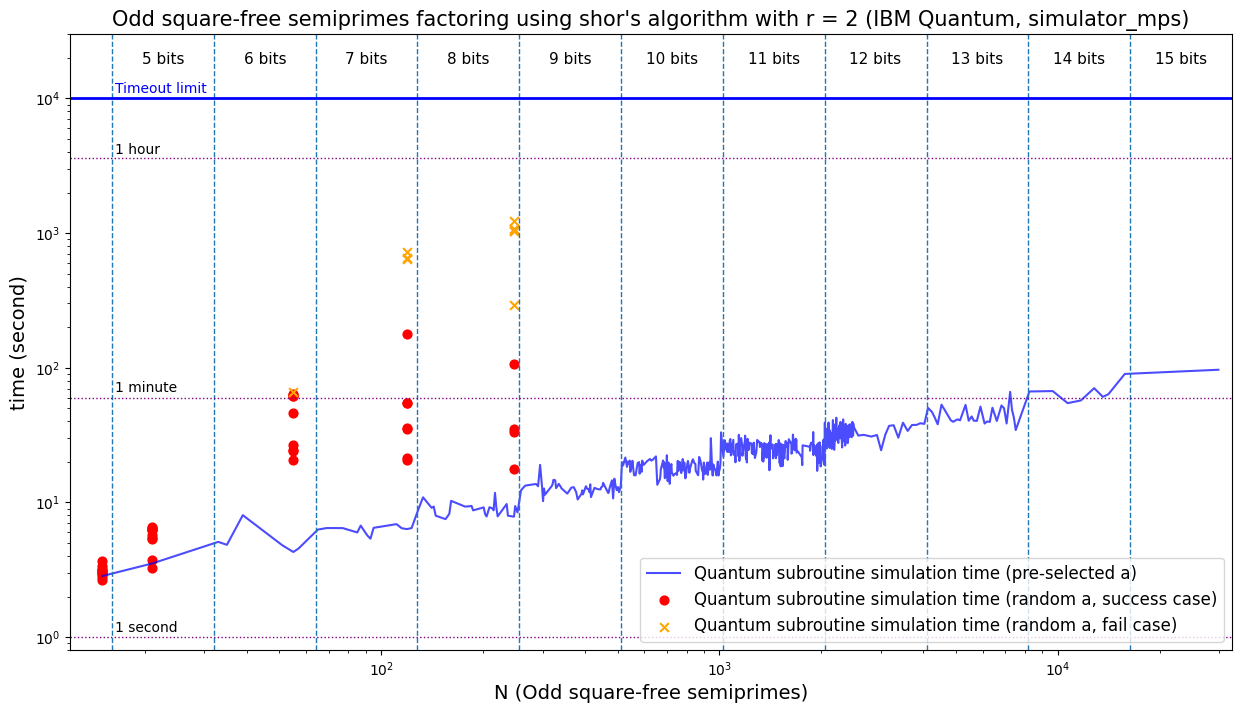}
\caption{The result of the quantum period-finding subroutine time}
\label{fig:fig3}
\end{figure}

Figure~\ref{fig:fig3} illustrates the performance of Shor's algorithm in terms of the required time for integer factorization given various inputs of $N$. The integer factorization time when parameter $a$ is pre-selected is depicted linearly, showing an overall tendency on a log-log scale. While Shor's algorithm theoretically exhibits polynomial time complexity with respect to $\log N$, the results obtained through classical simulation deviate from this prediction. The pattern remains generally consistent within the same number of bits, with a significant increase in the required time observed as the number of bits increases.

One noteworthy aspect of the graph is the scalability of performance measurement for integer factorization with respect to $N$. This implies that the measurements were not limited to specific numbers but were applicable to all possible values of $N$, enabling the prediction of the time required for integer factorization for any given $N$. This is highly significant as it allows for the prediction of the feasibility and time requirements of integer factorization for all possible values of $N$.

Furthermore, it is worth noting that all experimental results were obtained with a fixed number of shots (8 shots). When $a$ is pre-selected with a minimum period of $r=2$, 8 shots were sufficient to factorize all given input numbers.

On the other hand, in the case of random selection of $a$, the value of $r$ is not consistent, making it impossible to measure the factorization time consistently. The red circles on the graph demonstrate significant variations in the factorization time for the same number, depending on the trial, and the yellow X marks indicate instances of failure to produce a factorization result. (In this experiment, the timeout limit was set to 10,000 seconds.) Within this timeout limit, when pre-selection is performed, successful factorization was achieved up to 14 bits. However, when random selection was used, it was observed that successful factorization could not be achieved for numbers with 9 bits or more.

\section{Conclusion \& Future Works}
In this study, we evaluated the scalability of integer factorization performance using Shor's algorithm in quantum circuit simulation based on the Matrix Product State (MPS) method. Comparing the performance of integer factorization between pre-selection and random selection of parameters, we concluded that pre-selecting $a$ is beneficial for scaling quantum computing performance. Furthermore, it is crucial to be mindful of the impact of parameter $a$ on the performance of Shor's algorithm, as it determines the period of the function to be evaluated. This variable is expected to affect both the time required to generate quantum circuits and the simulation (computation) time of quantum circuits. Further research is needed to analyze the optimal parameter selection methodology through time complexity analysis for each step (quantum circuit generation, quantum circuit operation, continued fraction processing) in order to identify the most effective parameters.

Moreover, researchers have utilized the $\varepsilon$-random technique to optimize the number of quantum samples required for algorithms in Quantum Random Access Memory (QRAM)~\cite{12} based on Lemma~\ref{lem:levy}. Building on this technique, we intend to carry out theoretical research aimed to reduce the complexity of integer factorization and explore various factorization methodologies beyond Shor's algorithm.

\begin{lemma}[L\'{e}vy's inequality~\cite{13}]\label{lem:levy}
Let $f:S^{n}\rightarrow \mathbb{R}$ be a function defined on the $n$-dimensional hypersphere $S^n$ with Lipschitz constant
\begin{equation}
\eta=\sup_{\mathbf{x}_1, \mathbf{x}_2}\cfrac{\left| f(\mathbf{x}_1)-f(\mathbf{x}_2) \right|}{\| \mathbf{x}_1-\mathbf{x}_2 \|}<\infty,
\end{equation}
with respect to the euclidean norm $\| \cdot \|$ and a point $\mathbf{x}\in S^{n}$ be chosen uniformly at random. Then
\begin{equation}
\Pr\left[\left| f(\mathbf{x})-\mathbb{E}[f] \right| \ge \varepsilon \right] \le 2\exp\left(-\cfrac{C(n+1)\varepsilon^2}{\eta^2}\right),
\end{equation}
for some constant $C>0$.
\end{lemma}

\section*{Acknowledgements}
The author would like to express gratitude to Ju-Young Ryu for valuable discussions related to entanglement analysis and to Daun Jung for assistance with visualization issues. This work was partly supported by the National Research Foundation of Korea (NRF) through a grant funded by the Ministry of Science and ICT (NRF-2022M3H3A1098237).


\begin{thebibliography}{99}
\bibitem{1} Shor, P. W. (1994). Algorithms for quantum computation: discrete logarithms and factoring. In Proceedings 35th annual symposium on foundations of computer science (pp. 124-134). Ieee.
\bibitem{2} Grover, L. K. (1996). A fast quantum mechanical algorithm for database search. In Proceedings of the twenty-eighth annual ACM symposium on Theory of computing (pp. 212-219).
\bibitem{3} Shor, P. W. (1999). Polynomial-time algorithms for prime factorization and discrete logarithms on a quantum computer. SIAM review, 41(2), 303-332.
\bibitem{4} Bonnetain, X., Naya-Plasencia, M., \& Schrottenloher, A. (2019). Quantum security analysis of AES. IACR Transactions on Symmetric Cryptology, 2019(2), 55-93.
\bibitem{5} Jordan, S. P., \& Liu, Y. K. (2018). Quantum cryptanalysis: shor, grover, and beyond. IEEE Security \& Privacy, 16(5), 14-21.
\bibitem{6} Gerjuoy, E. (2005). Shor’s factoring algorithm and modern cryptography. An illustration of the capabilities inherent in quantum computers. American journal of physics, 73(6), 521-540.
\bibitem{7} Grassl, M., Langenberg, B., Roetteler, M., \& Steinwandt, R. (2016). Applying Grover’s algorithm to AES: quantum resource estimates. In International Workshop on Post-Quantum Cryptography (pp. 29-43). Cham: Springer International Publishing.
\bibitem{8} Rivest, R. L., Shamir, A., \& Adleman, L. (1978). A method for obtaining digital signatures and public-key cryptosystems. Communications of the ACM, 21(2), 120-126.
\bibitem{9} Dworkin, M. J., Barker, E. B., Nechvatal, J. R., Foti, J., Bassham, L. E., Roback, E., \& Dray Jr, J. F. (2001). Advanced encryption standard (AES).
\bibitem{10} Jiang, S., Britt, K. A., McCaskey, A. J., Humble, T. S., \& Kais, S. (2018). Quantum annealing for prime factorization. Scientific reports, 8(1), 17667.
\bibitem{11} Vidal, G. (2003). Efficient classical simulation of slightly entangled quantum computations. Physical review letters, 91(14), 147902.
\bibitem{12} Jeong, K. (2023). Sample-size-reduction of quantum states for the noisy linear problem. Annals of Physics, 449, 169215.
\bibitem{13} L\'{e}vy, P., \& Pellegrino, F. (1951). Probl\'{e}mes concrets d'analyse fonctionnelle.
\end{thebibliography}
\end{document}